\documentstyle[12pt]{article}
\def\bq{\begin{equation}}
\def\eq{\end{equation}}
\def\bqa{\begin{eqnarray}}
\def\eqa{\end{eqnarray}}
\def\roughly#1{\mathrel{\raise.3ex
    \hbox{$#1$\kern-.75em\lower1ex\hbox{$\sim$}}}}

\renewcommand{\thefootnote}{\fnsymbol{footnote}}
\begin{document}
\begin{flushright}
{BI-TP 2001/05}\\
\end{flushright}

\vspace{2 cm}

\begin{center}
{\large\bf On $e^+ e^- \rightarrow W^+ W^- \rightarrow 4 f (+\gamma)$ 
at LEP2}\footnote{Suppported by the BMFT, Bonn, Germany, Contract 05
HT9PBA2}
\end{center}

\renewcommand{\thefootnote}{\arabic{footnote}}
\setcounter{footnote}{0}
\vspace{1 cm}

\begin{center}
{\bf Yoshimasa Kurihara} \\
\vspace{0.2 cm}
  KEK, Tsukuba, Japan \\
\vspace{0.5 cm}
{\bf Masaaki Kuroda} \\
\vspace{0.2 cm}
Meijigakuin University, Yokohama, Japan \\
\vspace{0.5 cm}
{\bf Dieter Schildknecht} \\
\vspace{0.2 cm}
  Fakult\"at f\"ur Physik, Universit\"at
       Bielefeld\\
      Bielefeld, Germany
\end{center}

\vspace{2 cm}

\begin{flushleft}
Abstract
\end{flushleft}
The results on $e^+e^-\to W^+W^-\to 4f (+\gamma)$ obtained by
different groups are compared with each other.  Differences in the
results for the total cross section of up to about 0.6 \% 
are traced back to different ways  of
implementing the double-pole approximation.

\vfill\eject

\vspace{1cm}
In the workshop report on $W$-pair production at LEP2 \cite{lep2}, 
theoretical results on the cross section for the reaction $e^+ e^- \rightarrow
W^+ W^- \rightarrow 4 f (+ \gamma)$ from several working groups are presented. 
While the workshop has been in progress, based on our previous work \cite{KKS},
we obtained results for the total cross section of the above reaction 
that were
approximately 0.5 \% larger than the results published in ref.\cite{lep2}.
In the present note, we give our results, which remained unpublished so far, 
and we compare them with the results of the BBC \cite{BBC1} and the
RACOON \cite{DDRW1,DDRW2,DDRW3}
group in ref.\cite{lep2}.\footnote{Additional results, not published in 
refs.\cite{lep2, BBC1,DDRW2}, were kindly provided to us 
by the BBC group (Beenakker, 
Berends and Chapovsky) and the RACOON group (Denner, Dittmaier, Roth and 
Wackeroth).}

We will arrive at the conclusion that different ways of implementing the 
double-pole approximation (DPA) that enters the evaluation of the 
cross section may lead to differences up to about 0.6 \% in the total cross 
section for the above-mentioned reaction. 

We briefly summarize our approach \cite{KKS} that will be referred to as KKS 
in the subsequent discussion of the different approaches. 
The total cross section is written as the sum of the Born cross section, 
$\sigma^{{\rm full}}_B$, the virtual one-loop and soft-photon-radiation 
cross section, $\sigma_{V+S}$, and the hard-photon-radiation cross 
section, $\sigma_H$, 
\begin{equation} 
\sigma_{{\rm tot}} = \sigma^{{\rm full}}_B + \sigma_{V+S} + \sigma_H .
\end{equation}
As in ref.\cite{lep2}, the Born cross section, $\sigma^{{\rm full}}_B$, is 
based on the complete set of Feynman diagrams contributing to 
$e^+ e^- \rightarrow$ four fermions, i.e. diagrams not containing $W$-resonance
enhancements are also taken into account. 

The expression for $\sigma_{V+S}$ is given by
\bqa
   \sigma_{V+S}& =& {1\over {2s}} (2\pi)^{-7}{1\over{512}}
           {\sum} ^\prime \int  dk_+^2dk_-^2 d\cos\theta_W
              d\hat\varphi_2d\cos\hat\theta_2d\hat\varphi_4d\cos\hat\theta_4
        {{\vert \vec{k_\pm}\vert}\over {E_{beam}}}\label{5} \\
     && {{ \vert {\cal H }(e^+(\sigma_+)e^-(\sigma_-) \to
                 W^+_{\lambda_+}(f_1(\tau_1)\bar f_2(\bar\tau_2))
                 W^-_{\lambda_-}(f_4(\tau_4)\bar f_3(\bar\tau_3)))
           \vert^2 }\over{ [(k_+^2-M_W^2)^2+\Gamma_W^2 M_W^2]
               [(k_-^2-M_W^2)^2+\Gamma_W^2 M_W^2] }}
          \Big|_{\small{\cal O}(\alpha)},\nonumber
\eqa
where ${\sum}^\prime$ indicates the spin sum over the final fermions   
as well as the spin average of the initial $e^+e^-$ pair. 
The variables
\begin{equation}
k^2_+ , k^2_- , \cos\theta_W
\end{equation}
stand for the squared invariant masses and the cosine of the
polar production angle 
of the fermion pairs in the $e^+e^-$ center-of-mass frame, while 
\begin{equation}
\hat\varphi_2 , \cos \hat\theta_2; \hat\varphi_4 , \cos \hat\theta_4
\end{equation}
denote the {\it fermion angles in the rest frames} of the pairs 
of invariant masses 
$k^2_+$ and $k^2_-$, respectively; specifically, in the case of on-shell 
production, we have $k^2_+ = k^2_- = M^2_W$. As indicated by the integration 
variables in (2), the amplitude describing (on-shell or off-shell) 
$W$ production and decay in (2) is to be expressed  in terms of 
$k^2_+ , k^2_- , \cos \theta_W$ and the fermion-pair rest-frame angles 
$\hat\varphi_2 , \cos\hat\theta_2$ and $\hat\varphi_4 , \cos\hat\theta_4$. 
The choice of the rest-frame angles in (2), in view of the DPA,
is dictated by the requirement that the off-shell and on-shell decay 
configurations have identical (normalized)
decay probabilities.\footnote{The (normalized) 
{\it rest-frame decay distribution} is 
independent of the virtuality, i.e. independent of whether 
$k^2_{\pm} = M^2_W$ or $k^2_{\pm} \not= M^2_W$.} 
The DPA now consists of ignoring any additional 
dependence on $k_\pm^2$ in the $W$-boson (production and decay) 
amplitude in (2)
by evaluating the amplitude at the pole position, $k^2_+ = k^2_- = M^2_W$. 
The variation of the phase space with $k^2_\pm$ in (2) is taken 
into account via the 
dependence of the $W$-boson three momentum, $|\vec k_\pm |$, that is given 
by 
\bq
|\vec k_\pm |^2 = \frac{1}{4s} ( s^2 + k_+^4 + k_-^4 - 2s (k_+^2 + k_-^2) 
- 2 k_+^2 k_-^2),  \label{7}
\eq
where the $e^+ e^-$ center-of-mass energy squared is denoted by $s$. 

Also the hard-photon radiation is treated in the DPA, the expression for 
$\sigma_H$ being analogous to (2). 

The RACOON ansatz, when evaluating $\sigma_V$ in the DPA according to
(2),  differs from ours in their choice of the 
{\it laboratory-frame fermion angles}. 
This implies that in general off-shell and on-shell amplitudes at 
different rest-frame angles are associated with each other,
and accordingly an avoidable difference between the 
off-shell amplitude and its approximating on-shell one is introduced.
The choice of the laboratory angles also implies a dependence on which
pair of outgoing leptons (quarks) is used to specify the four-fermion
configuration.  
Fortunately, as noted by RACOON, using different pairs of fermions affects the 
result for the total cross section only at the 
level of 0.1\% \cite{DDRW3}.

In the treatment of the BBC group, also the 
phase space and the propagators in (2)
are evaluated on the $W$ mass shell, $k^2_\pm = M^2_W$, 
by carrying out the substitution
\begin{equation}
\frac{M_W \Gamma_W}{(k^2_\pm - M^2_W) + M^2_W \Gamma^2_W} 
\longrightarrow \pi \delta (k^2_\pm - M^2_W)
\end{equation}
that becomes exact in the zero-width limit\footnote{In ref.\cite{KKS},
such an approximation is called the zero-width approximation.} of $M_W 
\Gamma_W \rightarrow 0$.   As, by construction, there is no off-shell
phase-space element in the BBC ansatz, the question of which
on-shell amplitude is to be associated with a certain
off-shell phase-space element becomes meaningless.

Concerning the photon radiation, the RACOON treatment differs from 
the DPA ones of BBC and KKS by taking into account radiation from 
the full set of Born diagrams for $e^+e^-\to 4~ {\rm fermions}$. 

Finally, we note that RACOON in its treatment includes 
non-factorizable radiative corrections as well. 
Since such corrections cancel \cite{FKM,BBC2,DDRW4,CK}
in the total cross section, they were not included in the KKS 
evaluation.   In view of the discussion of differences
between the final results of the different groups below,
we note that RACOON did not explicitly verify the cancellation
of the non-factorizable radiative corrections in the sum of
$\sigma_V$, $\sigma_S$ and  $\sigma_H$. 
The non-factorizable radiative corrections taken into account by 
BBC, by construction, drop out in the total cross section.

We schematically summarize the assumptions underlying the approaches
by the different groups in Table 1.

\begin{table}
\begin{tabular}{|l| l|c|c|}\hline
  Group            & \multicolumn{2}{c|}{DPA} & photon radiation  \\
\cline{2-3}  
                   & \multicolumn{2}{c|}{phase space and fermion angles}   
                   &   \\
\hline
  BBC              & \multicolumn{2}{c|}{ zero-width approximation}
                   & DPA \\
\cline{2-3}  
                   &   on-shell & irrelevant  & \\
\hline
  KKS              & off-shell & $W^\pm$ rest frames & DPA \\ 
\hline
  RACOON           & off-shell & laboratory frame   & full    \\
\hline
\end{tabular}

\vspace{0.1 cm}
\noindent{\footnotesize Table 1. Comparison of the schemes of computing
radiative corrections by three groups.}
\end{table}

In our calculation, as in ref.\cite{lep2}, the $\alpha_{G_\mu}$ scheme is 
used, i.e. the coupling constants appearing at tree level are evaluated in 
terms of the muon decay constant, $G_\mu$. In order to make the comparison of
the results of the different groups transparent, we stick to comparing 
cross sections with order $\alpha$ corrections, discarding the inclusion of 
higher-order effects, such as the exponentiation of the initial state 
radiation, etc.

We use the set of input parameters also employed in the LEP2 workshop 
proceedings \cite{lep2}, 
\bqa
  M_Z&=&91.1867GeV,~~~~~M_W =80.35 GeV,~~~~ M_H = 150 GeV, \\ \nonumber
  G_\mu &=&  1.166389\times 10^{-5} GeV^{-2}, ~~
  \Gamma_W= 2.08699 GeV.
\eqa
The numerical integrations were performed by using the Monte Carlo 
program {\tt BASES} \cite{Kawabata}. 

Various checks of our program were performed:
\begin{itemize}
\item[i)]
The cancellation of the ultraviolet and the infrared divergences in the 
cross section was confirmed analytically as well as numerically. 
\item[ii)]
The stability against changing the soft-photon separation energy, $k_c$, was 
checked by varying $k_c$ between 0.001 GeV and 0.1 GeV. 
\item[iii)]
The cancellation of mass singularities in the total cross section was checked
by varying the fermion masses. 
\item[iv)]
The virtual correction, $\sigma_V$, and the soft photon radiation, $\sigma_S$,
were independently calculated by using {\tt GRACE} \cite{GRACE} as 
well as the program
constructed by one of the present authors \cite{MK}
several years ago. 
\end{itemize}

\noindent{\bf The Leptonic $W^\pm$ decay}\\
\indent  We start with the leptonic decay, 
$e^+ e^- \rightarrow W^+ W^-
\rightarrow (\bar\mu , \nu_\mu) ( \tau , \bar\nu_\tau ) (+\gamma)$. In 
Tables\footnote{We note that all 
cross sections displayed in Tables 2 to 8 are given in units of $fb$.}
2 to 4, our results (KKS) are compared with the ones from BBC and 
RACOON. In order to explicitly verify that our numerical evaluation is 
consistent with the one from BBC, we have also performed an 
evaluation
 \'a la BBC, i.e. by replacing the full phase space in (\ref{5}) 
by the zero-width approximation (\ref{7}). 
Compare KKS (zero-width) in Tables 2 to 4. 

\begin{table}[h]
\begin{tabular}{|c|l|l|l|}\hline 
     Group & $\sqrt s = 184$ GeV & $\sqrt s = 189$ GeV &$\sqrt s = 200$ GeV \\
\hline
   BBC            &           &            & 200.14    \\
   KKS(zero-width)& 181.84(9) &  190.67(9) & 200.14(9)    \\
   KKS            & 182.80(9) &  191.40(9) & 200.57(9) \\
   RACOON         & 182.53(11)&  190.96(11)& 199.98(12) \\
\hline
\end{tabular} 

\noindent {\footnotesize Table 2. The total cross section for 
the process $e^+e^-\to W^+W^-\to(\bar\mu \nu_\mu)(\tau \bar\nu_\tau)$ 
with strict order-alpha
corrections evaluated by BBC, KKS  and
RACOON at $\sqrt s=$184, 189 and 200 GeV.}

\begin{tabular}{|c|r|c|l||c|c|}\hline 
   Group &\multicolumn{1}{c|}{$\sigma_B^{\rm full}$} & $\sigma_{V+S+H}$ 
         &\multicolumn{1}{c||}{$\sigma_{tot}$} & $\sigma_V^{\rm fact}$ 
         & $\sigma_{V+S+H}$-$\sigma_V^{\rm fact}$  \\ \hline
   BBC             & 219.99(5) & -19.85 & 200.14 &      &   \\
   KKS(zero-width) & 219.85(7) & -19.71 & 200.14(9) & -437.92 & 418.21 \\ 
   KKS             & 219.85(7) & -19.28 & 200.57(9) & -422.87 & 403.59\\
   RACOON          & 220.06(6) & -20.08 & 199.98(12)& -422.75 & 402.67 
 \\ \hline
\end{tabular}

\noindent {\footnotesize Table 3. The full-Born cross section,
$\sigma_B^{\rm full}$, as well as the radiative correction, 
$\sigma_{V+S+H}$, adding up to the total cross section, 
$\sigma_{tot}$, 
for the process $e^+e^-\to W^+W^-\to
(\bar\mu \nu_\mu)(\tau \bar\nu_\tau)$ at  $\sqrt s=200$ GeV. 
The factorizable part of the virtual correction, 
$\sigma_V^{\rm fact}$, as well as the difference,
$\sigma_{V+S+H}$-$\sigma_V^{\rm fact}$, for  
$m_\gamma=10^{-6}$ GeV is also shown.}

\begin{tabular}{|c|l|c|c|c|}\hline 
   Group &  $\sigma_B^{\rm DPA}$ & $R_B^{\rm DPA}$ 
         & $\sigma_{V+S+H}$  & $R_{V+S+H}^{\rm DPA}$ \\ \hline 
   BBC             & 229.8(1)  & 3.5\% & -19.85 & 3.0\% \\
   KKS(zero-width) & 229.58(2) & 3.4\% & -19.71 & 2.2\%      \\
   KKS             & 221.72(4) &       & -19.28 &       \\
\hline
\end{tabular} 

\noindent {\footnotesize Table 4. The DPA-Born cross section,
$\sigma_B^{\rm DPA}$, as well as the order $\alpha$ correction 
due to virtual effects, soft and hard bremsstrahlung,
$\sigma_{V+S+H}$, for the process $e^+e^-\to W^+W^-\to
(\bar\mu \nu_\mu)(\tau \bar\nu_\tau)(+\gamma)$ at  $\sqrt s=200$ GeV,
and their relative changes as a result of replacing the off-shell
phase space by the zero-width approximation. 
The relative changes $R_B^{\rm DPA}$ and $R_{V+S+H}^{\rm DPA}$ are 
defined by
$R_B^{\rm DPA}=
           {{\sigma_B^{\rm DPA}(BBC)-\sigma_B^{\rm DPA}(KKS)}\over
             {\sigma_B^{\rm DPA}(BBC)}}$ etc.}
\end{table}
In the interpretation of the results in Tables 2 to 4, we proceed in several
steps: 
\begin{itemize}
\item[i)]
Comparing the results of KKS (zero-width) and BBC, in Tables 2 and 3, 
we find good agreement between  these independent calculations of 
$\sigma_{{\rm tot}}$ and the virtual correction $\sigma_{V+S+H}$. 
The comparison explicitly demonstrates that both calculations, BBC and KKS, 
are reliable, elementary errors being excluded. The relative 
difference between the BBC (the KKS(zero-width)) and the KKS results
is displayed 
in Table 4. Since the virtual corrections are proportional to the Born cross 
section evaluated in DPA, the relative change in $\sigma_{V+S+H}$ of about
3\% is approximately given by the relative change in $\sigma^{{\rm
DPA}}_B$. The relative change of around 3\% amounts to a decrease of the total 
cross section by about 0.3\% or 0.6 fb (compare Table 2) of 
KKS(zero-width) relative to KKS. 

We conclude that the 0.3\% reduction of $\sigma_{{\rm tot}}$ 
of BBC relative to KKS is due to the use of the simplifying zero-width
approximation that enlarges $\sigma_B^{\rm DPA}$ and accordingly
the absolute value of the (negative) radiative correction,
$\sigma_{V+S+H}$.
\item[ii)]
We turn to a discussion of the differences between the results of KKS and 
RACOON. According to Table 1, the increase of $0.6 fb \cong 0.3$\% 
in Tables 2 and 3 of $\sigma_{{\rm tot}}$ from KKS relative to $\sigma_{{\rm 
tot}}$ from RACOON  at $\sqrt s=200$ GeV is to be traced back to 
differences in the (soft and hard) 
photon radiation\footnote{The discussion is simplified, in so far as we
ignore the effect of the different choices of the fermion angles
and assume exact cancellation of the non-factorizable corrections
in the RACCON treatment.  Compare the more detailed discussion in the
case of hadronic $W^\pm$ decays below.}.
KKS treats photon radiation in the DPA, while RACOON radiates photons 
from all diagrams. 
Table 3 indeed shows that the virtual factorizable corrections, 
$\sigma^{{\rm fact}}_V$,
of KKS and RACOON agree very well, the differences in $\sigma_{V+S+H}$ 
and $\sigma_{{\rm tot}}$ thus 
indeed being due to the different treatment of the radiation. 
Compare the value of $\sigma_{V+S+H} - \sigma_V^{\rm fact}$ in Table 3. 
\item[iii)]
Finally, we compare the results of BBC and RACOON. 
As noted in Table 1, the BBC procedure differs from the one by
RACOON in applying the DPA to the radiation in conjunction with the use of
the zero-width approximation.  Since the zero-width approximation
decreases
the total cross section, the increase of the cross section of BBC relative to 
RACOON induced by
applying the DPA to the treatment of the radiation is 
(partially) cancelled.  The surprisingly good agreement
between $\sigma_{tot}$ by RACOON and by BBC at 200 GeV appears as an accidental
one. 
In this connection it is to be noted that the difference between RACOON and 
BBC, or rather KKS (zero-width), according to Table 2, changes with decreasing 
$e^+ e^-$ energy from $+0.16 fb$ to $-0.7 fb$. 
\end{itemize}

\vspace {0.2 cm}

\noindent{\bf The Hadronic $W^\pm$ decay}\\
\indent   For the hadronic channel, $e^+e^-\to W^+W^- \to (u \bar d)(\bar c s)$,
no results from BBC are  available. We compare our
results, KKS and KKS(zero-width), with the result
from RACOON.

In our calculations, we used values of the quark masses that coincide
with the ones from RACOON (in units of GeV/c$^2$),
\bqa
    m_u&=& 0.0485,~~~ m_c = 1.55,~~~  m_t = 174.17 , \\ \nonumber
    m_d&=& 0.0485,~~~ m_s = 0.15, ~~~m_b = 4.5.
\eqa
The quark masses enter $\sigma_V$, $\sigma_S$ and $\sigma_H$ in terms
of mass-singular logarithms, but are obviously irrelevant for the total
cross section. \\
\indent    
\begin{table}[h!]
\begin{tabular}{|c|c|c|c|}\hline
  Group & $\sqrt s= 184$ GeV & $\sqrt s= 189$ GeV  
        & $\sqrt s= 200$ GeV \\
\hline
   KKS({\rm zero-width}) & 1636.9(9) & 1716.0(9) &  1801.5(9) \\
   KKS      &  1647.5(9) & 1724.7(9) & 1807.6(9)\\    
   RACOON &  1640.7(7) & 1716.4(8) & 1797.4(9)\\
\hline
\end{tabular}
\medskip

\noindent{\footnotesize Table 5.  The comparison of the total cross section
for the process $e^+e^- \to (u \bar d)(\bar c s)(+\gamma)$ at
$\sqrt s=184, 189$ and 200 GeV.  The RACOON results are taken from
ref.\cite{DDRW2}.}

\begin{tabular}{|c|c|c|c||c|c|}\hline 
   Group &  $\sigma_B^{\rm full}$ & $\sigma_{V+S+H}$ 
         & $\sigma_{tot}$ & $\sigma_V^{\rm fact}$ 
         & $\sigma_{V+S+H}$-$\sigma_V^{\rm fact}$  \\ \hline
   KKS(zero-width) & 1980.4(4) & -178.9 & 1801.5 & -3441.7 & 3261.6 \\ 
   KKS             & 1980.4(4) & -172.8 & 1807.6 & -3323.2 & 3150.4\\
   RACOON          & 1980.8(7) & -183.4 & 1797.4 & -3318.2 & 3134.8 
 \\ \hline
\end{tabular} 

\noindent {\footnotesize Table 6. The full-Born cross section,
$\sigma_B^{\rm full}$, as well as the radiative correction, 
$\sigma_{V+S+H}$, adding up to the total cross section, 
$\sigma_{tot}$, 
for the process $e^+e^-\to W^+W^-\to
(u, \bar d)(\bar c s)(+\gamma)$ at  $\sqrt s=200$ GeV. 
The factorizable part of the virtual correction, 
$\sigma_V^{\rm fact}$, as well as the difference,
$\sigma_{V+S+H}$-$\sigma_V^{\rm fact}$, for 
$m_\gamma=10^{-6}$ GeV is also shown.}

{\small
\begin{tabular}{|c|c|l|c|c|l|l|c|}\hline
  Group   & $\sigma_B^{\rm DPA}$
          & \multicolumn{2}{c|}{$\sigma_V$}  
          & \multicolumn{2}{c|}{$\sigma_S(E_\gamma<0.1)$} 
          & \multicolumn{2}{c|}{$\sigma_H(E_\gamma>0.1)$} \\ \cline{3-8}
          &  & fact. & non-fact. & fact. & non-fact. 
          & fact. & non-fact.    \\  
\hline
  KKS     & 1994.7(1)  
          &-3323.2(4) &    & 1695.5(3)  & 46.2  & 1454.4 &  \\ \cline{3-8}
          & &\multicolumn{2}{c|}{}
          &\multicolumn{2}{c|}{1741.7(3)} &\multicolumn{2}{c|}{}\\
\hline
 KKS  & 1991.7(3) & -3317.3(6)  &   &   & &    &  \\ \cline{3-8}
 (lab.)    & & \multicolumn{2}{c|}{} &\multicolumn{2}{c|}{}
           &\multicolumn{2}{c|}{}\\
\hline
  RACOON   & 1992.2(7)  & -3318.2  & -59.2 &   &  &  &   \\ \cline{3-8}
           &     &\multicolumn{2}{c|}{-3377.4}
                 &\multicolumn{2}{c|}{1739.2} 
                 &\multicolumn{2}{c|}{1454.8} \\
\hline
\end{tabular}
} 
\vspace {0.05 cm}
{\footnotesize \noindent Table 7.  Comparison of $\sigma_V$, $\sigma_S$
and $\sigma_H$
for $e^+e^-\to W^+W^- \to (u \bar d)(\bar d s)$  
at $\sqrt s=200$ GeV for  $k_c=0.1$ GeV and for $m_\gamma=10^{-6}$
GeV. (Note that the slight difference between the sum, $\sigma_{V+S+H}$,
calculated from the results in this Table, and $\sigma_{V+S+H}$ in Table 6, is
due to the somewhat larger value of $k_c$).}
\end{table}

In Tables 5 to 7, our results are compared with the ones from
RACOON.  According to Tables 5 and 6, the observed pattern 
of deviations
between KKS, KKS(zero-width) and RACOON is similar to the one
found in the case of the leptonic $W^\pm$ decays:

\newpage

\begin{itemize}
\item[i)]  The results for $\sigma_{tot}$ from KKS at $\sqrt s=200$ GeV lie
about 10 $fb$ or 0.55\% above the ones from RACOON\footnote{In \cite{lep2}, 
the results from RACOON are also compared with the results from YFSWW, 
by Jadach et al., whereby also the dominant higher-order radiative corrections
are included. The results by YFSWW lie up to about 0.4\% above the ones from 
RACOON.}.

\item[ii)]  The results from KKS(zero-width), which simulate the zero-width
treatment employed by BBC, are only 3 $fb$ or 0.17\% larger than the
ones from RACOON.  As discussed under i) and iii) for the leptonic
$W^\pm$ decay, the zero-width approximation decreases the total
cross section, thus leading to an (accidental) coincidence of the
RACOON and KKS(zero-width) result.
\end{itemize}
   Remembering Table 1, the difference between KKS and RACOON
is  expected to be due to the difference in the treatment of the
photon radiation (DPA versus radiation from the full set of 
Born diagrams) and to the different choice of the fermion angles
kept invariant under the on-shell continuation in $k_\pm^2$.
In addition, we have to keep in mind that RACOON, as a consequence
of the complexity of their program,
did not explicitly verify the cancellation of the non-factorizable
corrections in their calculation of $\sigma_{V+S+H}$. \\
\indent  In the ensuing interpretation of the difference between RACOON
and KKS, we proceed in two steps:
\begin{itemize}
\item[i)]  We assume that the result by RACOON, even though being
obtained
by including the non-factorizable corrections to 
$\sigma_V$, $\sigma_S$ and $\sigma_H$ , implicitly contains
a cancellation of these contributions in the sum, $\sigma_{V+S+H}$.
The last column of Table 6 then shows that the factorizable  
contribution to photon radiation, 
$\sigma_{S+H}^{\rm fact}$, by KKS exceeds the one by RACOON by
15 $fb$. The difference in $\sigma_V^{\rm fact}$ amounts to
-5$fb$. Altogether  this adds up to the mentioned difference of 
10 $fb$ in $\sigma_{tot}$.\\
In order to clarify the origin of the difference of 5 $fb$ in
$\sigma_V^{\rm fact}$, between KKS and RACOON, we look at
Table 7.  In Table 7, in addition to the results from KKS and
RACOON, we provide results (i.e. KKS (lab.))
for $\sigma_B^{\rm DPA}$ and $\sigma^{{\rm fact}}_V$ employing laboratory 
fermion
angles, as used by RACOON in their treatment of $\sigma_V$.  
The use of the laboratory-frame fermion angles decreases the 
KKS results for $\sigma_V^{{\rm fact}}$ 
by 0.15 \% to agree with the ones from RACOON.
The difference of 5 $fb$ in $\sigma_V^{\rm fact}$ is accordingly
traced back to the different choice of the fermion angles kept
constant under extrapolation in $k_\pm^2$\footnote{The absolute value of 
$\sigma_V^{{\rm fact}}$ being dependent on the choice of the regularizing
photon mass, the comparison of $\sigma_V^{{\rm fact}}$ of KKS and RACOON is 
carried out for technical reasons and is meant to explicitly demonstrate the 
consistency of the results under identical boundary conditions.}.

\item[ii)]  
Alternatively, let us assume that the different treatment of the 
photon
radiation by KKS and by RACOON nevertheless leads to identical results
for the factorizable part of the photon radiation,
i.e. we assume that the factorizable part of the photon radiation,
even though unknown to the RACOON treatment, can be well
represented by the KKS results.
For $\sigma_S$, such a hypothesis might be suggested 
by the excellent agreement of KKS and RACOON  for $\sigma_S$
given in Table 7.
Under this assumption, the non-factorizable part of the 
radiative corrections, implicitly contained in the RACOON result,
can be calculated from the difference of the KKS and RACOON
results for $\sigma_V$, $\sigma_S$ and $\sigma_H$. 
According to Table 7, it becomes
\bqa
    \sigma_{V+S+H}^{\rm non-fact}&=& \sigma_V^{\rm non-fact}
       + \sigma_S^{{\rm non-fact}}
       +\bigl( \sigma_H^{\small \rm RACOON}-\sigma_H^{\rm fact}\bigr)
        \nonumber \\
      &=& -12.6 fb.
\eqa
Under the above hypothesis that the magnitude of the factorizable
part of the photon radiation is unaffected by replacing the full set of 
diagrams by the reduced set of the DPA,
we thus arrive at the conclusion that the RACOON
result for $\sigma_{V+S+H}$ contains a non-factorizable
contribution to the photon radiation that is not cancelling the
non-factorizable contribution to $\sigma_V$.
This non-factorizable contribution of about -12 $fb$ 
would then be responsible for the 
difference in $\sigma_{tot}$ of about $- 10 fb$ between KKS and RACOON.
\end{itemize}

   Actually, both the hypotheses discussed in i) and ii) may be
considered as extreme cases.  A detailed analysis of the
RACOON calculation may indeed reveal that both, the different  
treatment of the radiation as well as some mismatch of
non-factorizable corrections, are  at the root of the 0.6\% 
difference of the results of KKS and RACOON on $\sigma_{tot}$.

\vspace{0.2 cm}

\noindent{\bf Conclusions}\\
\indent By explicitly simulating the results of BBC and
RACOON, whenever possible, we found excellent agreement with BBC 
and RACOON.  Due to the complexity of the calculations, this in itself
is a remarkable result. Differences among the final results
up to about 0.6 \% were traced back to the different underlying
implementations of the double-pole approximation.
Roughly speaking, the different treatment of the photon
radiation (by using the DPA) leads to an increase of our results
relative to the ones of RACOON.
This increase is (partially) cancelled, 
if the additional simplifying approximation of on-shell kinematics (the BBC
zero-width approximation) is adopted.   
As a consequence of the known difficulties
concerning the treatment of unstable particles, it remains
an open question, 
to what extent even a technically extremely demanding 
full-one-loop evaluation
(without DPA) could possibly reduce theoretical uncertainties.

\vspace{1 cm}

\newpage

\noindent{\bf Acknowledgement}\par
    We would like to thank S. Dittmaier and S. Chapovsky 
for communicating their numerical results to us, and for providing details on
the structure of their computational procedures. 
We would like to also thank J. Fujimoto and T.Ishikawa
for valuable discussions and suggestions.

\vfill\eject
\end{document}